\documentclass[twocolumn,trackchanges]{aastex63}
\RequirePackage{lineno}
\usepackage{amsmath}
\usepackage{amssymb}
\usepackage{amsthm}
\usepackage{natbib,aasdefs,url,bm}
\usepackage{array}
\usepackage{float}
\usepackage{graphicx}
\usepackage{subfigure}
\usepackage{color}
\usepackage{threeparttable} % to use table notes
\usepackage{CJK}
\usepackage{lineno}
\usepackage{tablefootnote}
\usepackage{threeparttable}

\shorttitle{Another Neutrino-Coincident TDE Candidate with a Dust Echo?}
\shortauthors{Yuan, Winter and Lunardini}

\begin{document}
\begin{CJK*}{UTF8}{gbsn}

\title{AT2021lwx: Another Neutrino-Coincident Tidal Disruption Event with a Strong Dust Echo?}

\correspondingauthor{Chengchao Yuan}
\email{chengchao.yuan@desy.de}

%\correspondingauthor{Walter Winter}
%\email{walter.winter@desy.de}

\author[0000-0003-0327-6136]{Chengchao Yuan (袁成超)}\affil{Deutsches Elektronen-Synchrotron DESY, 
Platanenallee 6, 15738 Zeuthen, Germany}

\author[0000-0001-7062-0289]{Walter Winter}\affil{Deutsches Elektronen-Synchrotron DESY, 
Platanenallee 6, 15738 Zeuthen, Germany}

\author[0000-0002-9253-1663]{Cecilia Lunardini}\affil{Department of Physics, Arizona State University, 450 E. Tyler Mall, Tempe, AZ 85287-1504 USA}

\begin{abstract}
We discuss the possible association of an astrophysical neutrino (IC220405B) with the recently reported, extremely energetic tidal disruption event (TDE) candidate AT2021lwx (ZTF20abrbeie, aka ``Scary Barbie'') at redshift $z=0.995$. Although the TDE is about $2.6^\circ$ off the direction of the reconstructed neutrino event (\emph{outside} the 90\% {confidence level} localization region), the TDE candidate shares some important characteristics with so far reported neutrino-TDE associations: a strong {infrared} dust echo, high bolometric luminosity, a neutrino time delay with respect to the peak mass accretion rate of the order of hundred days, and a high observed X-ray luminosity. {We interpret this new association using an isotropic emission model, where neutrinos are produced by the collision of accelerated protons with infrared photons. After accounting for the high redshift of AT2021lwx (by interpreting the data in the SMBH frame), we find that} 
the expected neutrino fluences and neutrino time delays  are qualitatively comparable to the other TDEs.
Since data are only available up to 300 days post-peak in the SMBH frame, significant uncertainties exist in the dust echo interpretation, and therefore in the predicted number of neutrinos detected, $\mathcal N_{\nu}\simeq3.0\times10^{-3}-0.012$. We recommend further follow-up on this object for an extended period and suggest refining the reconstruction of the neutrino arrival direction in this particular case.
\end{abstract}

%ApJL limits: Abstract – no more than 250 words; Main Text – no more than 3500 words (not including acknowledgments, appendices or other supplementary material); Figures and Tables – no more than 5 combined figures (each limited to 9 panels) and tables, e.g. 3 figures and 2 tables.

%%%%%%%%%%%%%% Section 1: Introduction %%%%%%%%%%%%%
\section{Introduction}

Tidal disruption events (TDEs) are energetic optical transients that originate from stars that are tidally destroyed as they transit within the tidal radius of a supermassive black hole (SMBH). Approximately half of the stellar mass remains bound, and its subsequent accretion could power an electromagnetic flare that lasts from months to years \citep{1988Natur.333..523R,1989IAUS..136..543P}. Nowadays, increasingly detailed multi-wavelength observations of TDEs by the Zwicky Transient Facility \citep[ZTF,][]{2019PASP..131a8002B}, the  Wide-field Infrared Survey Explorer \citep[{WISE},][]{2010AJ....140.1868W}, and X-ray/radio surveys, such as eROSITA \citep{2021MNRAS.508.3820S} and Very Large Array Sky Survey \citep[VLASS,][]{2020PASP..132c5001L}, facilitate comprehensive modeling of the radiation processes and cosmological distributions of this source population \citep[e.g.,][]{2021ApJ...908....4V,2023ApJ...942....9H,2023ApJ...955L...6Y}. 

{Among the hundred or so observed TDEs and TDE candidates, three have been found to be coincident -- in time and position -- with three IceCube astrophysical neutrino events. {They include 
an event that was classified as a TDE with high confidence, AT2019dsg \citep{2021NatAs...5..510S},  and two TDE candidates, AT2019fdr \citep{2022PhRvL.128v1101R} and AT2019aalc \citep{2021arXiv211109391V} (associated neutrinos: IC191001A, IC200530A and IC19119A, respectively).}
These three candidate neutrino emitters}
%In the TDE catalog, one ZTF-identified TDE, AT2019dsg \citep{2021NatAs...5..510S}, and two TDE candidates, AT2019fdr \citep{2022PhRvL.128v1101R} and AT2019aalc \citep{2021arXiv211109391V}, are potentially correlated with three IceCube astrophysical neutrino events: IC191001A, IC200530A and IC19119A, respectively. 
 share some prominent similarities. For instance, they all exhibit high optical/ultra-violet (OUV) luminosities accompanied by bright and delayed infrared (IR) emissions, which have been interpreted as dust echoes, i.e.,  reprocessed radiation from the OUV and X-ray bands into IR wavelengths by surrounding dust 
\citep{2016MNRAS.458..575L,2016ApJ...828L..14J,2016ApJ...829...19V}. These three TDEs are located within the 90\% C.L. localization region of the corresponding neutrino events \citep{2023arXiv230401174A}, with an angular deviation $\Delta\theta\sim1.3^\circ-1.9^\circ$ \citep{2021arXiv211109391V}. The neutrino events were detected with significant time delays -- approximately 150-300 days -- after the OUV peak in the observer's frame. 
 %, known as the neutrino time delays. 

{These three neutrino-TDE associations identify TDEs as potential cosmic ray accelerators, as neutrinos are a byproduct of hadronic processes.}
%The neutrino counterparts of these three TDEs emphasize the acceleration of cosmic-ray primaries
%since the associated neutrinos are primarily generated in hadronic processes.
Many models including relativistic jets \citep{2011PhRvD..84h1301W,2016PhRvD..93h3005W,2017MNRAS.469.1354D,2017PhRvD..95l3001L,2017ApJ...838....3S}, accretion disks \citep{2019ApJ...886..114H}, wide-angle outflows/hidden winds \citep{2020ApJ...904....4F}, and tidal stream interactions \citep{2015ApJ...812L..39D,2019ApJ...886..114H}  have been proposed as the origin of the non-thermal electromagnetic (EM) and neutrino emission from TDEs, which could potentially explain these TDE-neutrino coincidences. {Some models also offer a physical explanation of the observed neutrino time delays} \citep{2020PhRvD.102h3028L, 2021NatAs...5..472W, 2022MNRAS.514.4406W, 2020ApJ...902..108M, 2021NatAs...5..436H,2023ApJ...948...42W,2023arXiv230902275M}. In the context of multi-messenger astrophysics, TDE parameters and neutrino detectability can be constrained using X-ray or $\gamma$-ray {upper bounds} \citep[see, e.g.,][for the application to AT2019dsg and AT2019fdr]{2023ApJ...956...30Y}. 

Recently, two additional dust-obscured TDE candidates, exhibiting strong dust echo signatures, were reported to be spatially and temporally coincident with Gold-type {(the chance of astrophysical origin is larger than $50$\%)}  astrophysical neutrinos events at IceCube \citep{2023ApJ...953L..12J}. %For these, the OUV emissions are nearly negligible compared to the former three, and the angular deviations between the reconstructed neutrino direction and the TDEs are also higher, $\Delta\theta\sim2.4^\circ-2.7^\circ$, although still within the 90\% localization region. 

{Here, we point out the potential coincidence between another energetic TDE candidate, AT2021lwx (ZTF20abrbeie, aka ``Scary Barbie"), and an IceCube neutrino event, IC220405B \citep{2022GCN.31842....1N}. Noting that AT2021lwx is not within the 90\% confidence level (C.L.) area of the neutrino event, and IC220405B is classified as one Bronze-type neutrino alert (where the chance of astrophysical origin is larger than $30\%$), the probability of the neutrino-correlation of AT2021lwx is lower than the candidates mentioned before. However, AT2021lwx shares some prominent signatures with other neutrino-coincident TDEs and candidates, encompassing a strong dust echo which explains the IR observations, high bolometric OUV and X-ray luminosities, and a comparable time delay of the neutrino detection. These similarities suggest that AT2021lwx may be another member of a class of neutrino-emitting TDEs, for which a common underlying mechanism exists. In this context, it is interesting to investigate the neutrino correlations and multi-messenger implications.}

{In this paper, we offer an interpretation of AT2021lwx in terms of a model where neutrinos and EM cascade emissions originate from accelerated protons colliding with IR target photons, similar to the one presented in \cite{2023ApJ...948...42W} and \cite{2023ApJ...956...30Y} (model ``M-IR'' therein). We first discuss the likelihood of the neutrno correlation in Sec. \ref{subsec:neu_corr}.
Given the crucial role of the IR radiation, }in Sec. \ref{subsec:dust_echo},
{we fit the IR light curve using early-time (ET) component in addition to a delayed component produced by the spherical dust torus}, and discuss the uncertainties arising from the absence of late-time IR data. In Sec. \ref{sec:Neu_em}, we further employ an isotropic wind model, constructed based on OUV/IR/X-ray observations, as described in \cite{2023ApJ...948...42W} and \cite{2023ApJ...956...30Y}, to investigate the spectral and temporal signatures of neutrino and EM cascade emissions produced within the dust radius. 
%In this treatment, the photomeson ($p\gamma$) interactions between thermal OUV/IR/X-ray photons and injected protons would dominate the neutrino production and subsequent trigger EM cascades. 
In addition, we compare AT2021lwx to AT2019dsg/fdr/aalc in terms of bolometric OUV and IR luminosities, as well as predicted neutrino fluences, and discuss the $\gamma$-ray constraints and the {likelihood of producing one neutrino event at IceCube in Sec. \ref{sec:dission}.}
%neutrino detectability. %

\section{AT2021lwx}\label{sec:AT2021lwx}
\subsection{{Localization of AT2021lwx and IC220405B}}\label{subsec:neu_corr}

Motivated by the similarities among the neutrino-coincident TDE candidates, {here we investigate} the potential spatial and temporal coincidence between AT2021lwx and IC220405B \citep{2022GCN.31842....1N}. AT2021lwx was initially discovered by ZTF on the 13th of April 2021 and was classified as a TDE candidate at redshift of $z=0.995$  \citep{2023ApJ...948L..19S}. The peak bolometric optical luminosity, after correcting for extinction, is exceptionally high, reaching, $10^{46}\rm ~erg~s^{-1}$. Multi-wavelength follow-ups have revealed bright X-ray and IR emissions \citep{2023MNRAS.522.3992W}. The latter has been preliminarily interpreted as a dust echo. We point out one {Bronze-type} neutrino alert, IC220405B\footnote{{This neutrino event was initially named as IC220405A \citep{2022GCN.31839....1I}}}, which is close to the TDE direction with an angular offset of $\Delta\theta\simeq2.6^\circ$ degrees, and arrived approximately 370 days after the OUV peak, equivalent to $\sim185$ days in the SMBH rest frame. 

\begin{figure}
    \centering
    \includegraphics[width=0.49\textwidth]{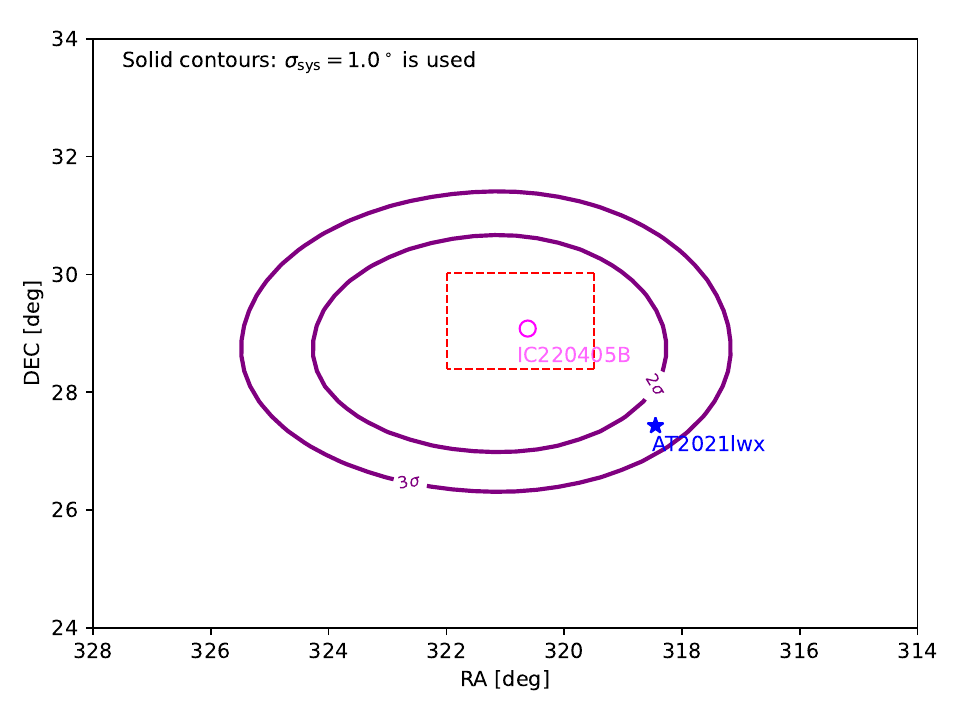}
    \caption{Localization of IC220405B and AT2021lwx. The red dashed rectangular shows the 90\% C.L. containment region of IC220405B \citep{2022GCN.31842....1N}. The inner and outter purple contours correspond to the $2\sigma$ and $3\sigma$ containment areas obtained by taking into account the systematic uncertainty $\sigma_{\rm sys}=1.0^\circ$. }
    \label{fig:neu_loc}
\end{figure}

{Fig. \ref{fig:neu_loc} depicts the locations of AT2021lwx and IC220405B. We obtain the $2\sigma$ and $3\sigma$ containment areas of the neutrino event by performing the Gaussian extrapolation of the 90\% C.L. box\footnote{{More specifically, we derive the Gaussian standard deviations for each side of the rectangle using the corresponding 90\% C.L. uncertainties, and then employ a 2-dimensional Gaussian distribution to obtain the contours. Note that
the position error is statistical only; there is no systematic added.}} \citep[the dashed red rectangular,][]{2022GCN.31842....1N} and applying a systematic uncertainty of arrival directions $\sigma_{\rm sys}=1.0^\circ$ motivated by the estimates in \cite{2013Sci...342E...1I} and \cite{2020ApJ...894..101P}.} 

{We find that AT2021lwx locates in the $3\sigma$ containment contour in the refined localization analysis. 
Caution must be exercised when establishing the significance of the association since it is inferred from the initial 90\% C.L. localization box, and a more precise localization constraint requires detailed point source reconstructions from the IceCube Collaboration.}

\subsection{Dust echo modeling}\label{subsec:dust_echo}

Modeling of ZTF photometry indicates that AT2021lwx was produced by the tidal disruption of a massive star of $M_\star\sim14~M_\odot$ by a SMBH of $M_{\rm BH}\sim10^8~M_\odot$ \citep{2023ApJ...948L..19S}. However, it is important to note that this object is not exclusively identified as a TDE, given the low likelihood of such an event involving a massive star. At first, it was classified as a flare from an active galactic nucleus \citep[AGN,][]{2022TNSAN.195....1G};  another plausible interpretation is an {unusually powerful} accretion of a giant molecular cloud by a SMBH of $10^8-10^9~M_\odot$ \citep{2023MNRAS.522.3992W}. Nevertheless, the classification does not significantly influence our multi-messenger modeling, since our model ingredients, such as the proton injection rate and target photon fields, are build on the OUV/IR/X-ray observations, including the light curves and the spectra.

%We will show in this section that the geometry, such as the radius of the dust torus, can be determined by the shape of the IR lightcurve. 

\begin{figure}[htp]
    \centering
    \includegraphics[width=0.49\textwidth]{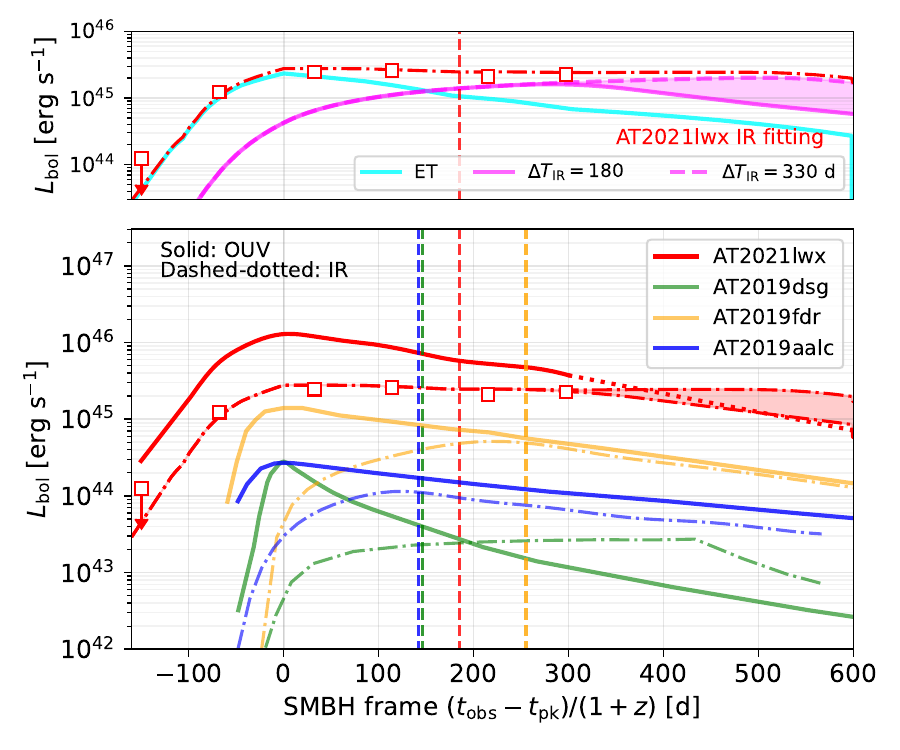}
    \caption{{\emph{Upper panel}:} IR light curve interpretation for AT2021lwx in the SMBH frame. The IR bolometric luminosity derived from {WISE W1/W2} light curves \citep{2023MNRAS.522.3992W} are shown as the square dots. {The early-time (ET) and spherical dust torus components} are depicted respectively as the cyan and magenta curves, whereas the red dashed-dotted curve is the total IR luminosity. The magenta area corresponding to the uncertainties of the IR time delay, e.g., $\Delta T_{\rm IR}=180-330$ d. {\emph{Lower panel}:} {bolometric} OUV (solid curves) and IR (dashed-dotted curves) light curves of AT2021lwx and the other three neutrino-coincident TDEs \citep{2023ApJ...948...42W} in SMBH frame. The vertical dashed lines show the detection times of the corresponding neutrino events.} %In both cases, the times are measured with reference to the OUV peak time ($t_{\rm pk}$).}
    \label{fig:4TDEs}
\end{figure}

\begin{figure}[htp]
    \centering
    \includegraphics[width=0.45\textwidth]{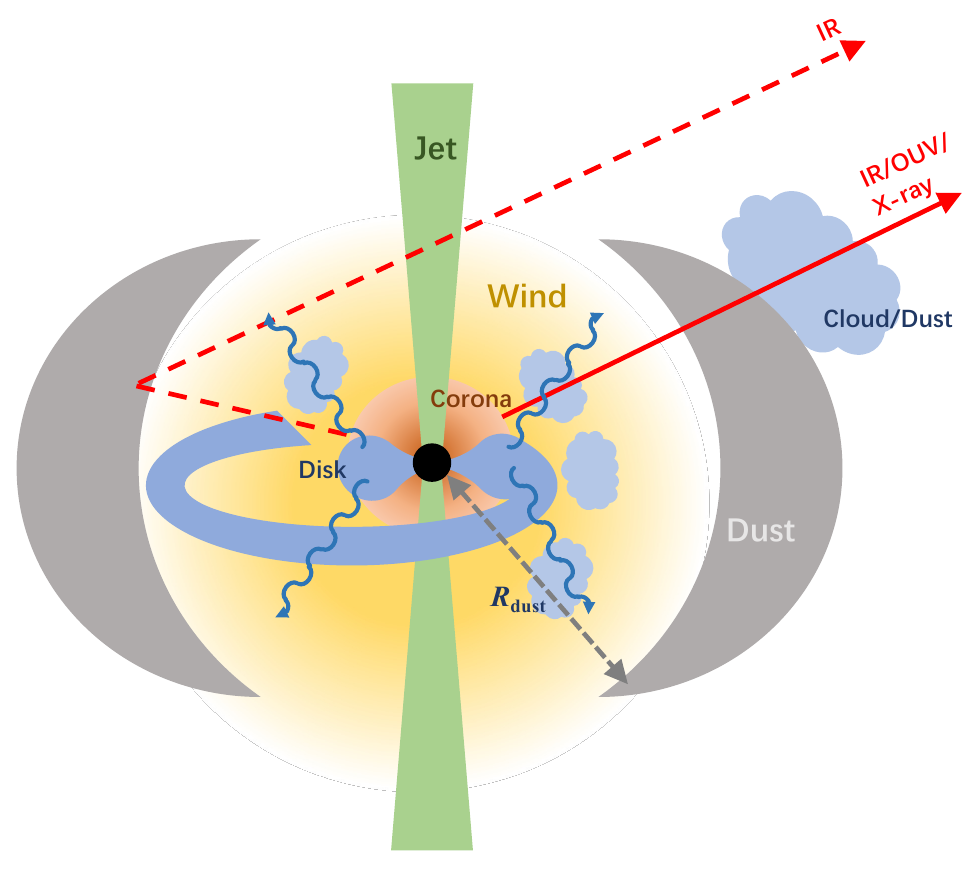}
    \caption{Schematic picture of the dust echo model. The central SMBH, accretion stream, accretion disk, disk corona, wind, dust torus and potentially a jet are shown. To fit the IR light curve, we model the dust torus as a spherical segment (e.g., the dashed red lines) and consider an additional early-time component (e.g., the red solid line). {The physical origins of the ET component are discussed in the main text.} The radius of dust torus $R_{\rm dust}$ determines the IR time delay.}
    \label{fig:schematic}
\end{figure}

For AT2021lwx, the IR light curve was measured by WISE in {W1} and {W2} bands before and after the OUV peak. {The IR data points in the upper panel of Fig. \ref{fig:4TDEs} and the OUV light curve of AT2021lwx are taken from \cite{2023ApJ...948L..19S} and \cite{2023MNRAS.522.3992W}, respectively, following bolometric corrections.} The OUV and IR spectra are consistent with black body distributions and the temperatures are measured to be $T_{\rm OUV}\sim1.2-1.6\times10^4$ K (1.03$-$1.38 eV) and $T_{\rm IR}\sim10^3$ K (0.9 eV) in the SMBH rest frame \citep{2023ApJ...948L..19S,2023MNRAS.522.3992W}. Using these temperatures, we calculate the bolometric correction factors, defined as the ratio of the energy flux from the entire black body spectrum to the energy flux in the r/g/{W1}/{W2} bands. From that, we {then obtain the OUV and IR bolometric luminosities, denoted as $L_{\rm OUV}$ and $L_{\rm IR}$, respectively. We stress that here $L_{\rm OUV}$ is corrected for extinction; as will be clear from the derivation in the reminder of this section.} The red square markers in Fig. \ref{fig:4TDEs} show  {$L_{\rm IR}$  as  inferred from WISE measurements, as a function of the time in SMBH rest frame}, e.g., $(t_{\rm obs}-t_{\rm pk})/(1+z)$, where $t_{\rm obs}$ is the time in the observer's frame and $t_{\rm pk}$ is the time that OUV luminosity peaks. {The red solid curve in the lower panel represents $L_{\rm OUV}$ for the time interval where data exist; its extrapolation to later times is shown as a red dotted curve. We find the peak values of $L_{\rm OUV}$ and $L_{\rm IR}$ to be $\simeq1.2\times10^{46}~\rm erg~s^{-1}$ and $\simeq3.1\times10^{45}~\rm erg~s^{-1}$, respectively. Our bolometric OUV luminosity, $L_{\rm OUV}$, is roughly a factor of $\sim$2 higher than the value in \cite{2023ApJ...948L..19S}, {since we corrected it for absorption by ambient dust, which induces the dust echo. This correction will be introduced in the end of this section}.}

The first impression we obtain from Fig. \ref{fig:4TDEs} is that AT2021lwx exhibits a comparable neutrino time delay with the other three TDEs in the SMBH frame (see the vertical lines in Fig. \ref{fig:4TDEs}), and its flat/steady IR luminosity after the OUV peak is consistent with a dust echo.  This dust echo interpretation is supported by the measured IR temperature, $T_{\rm IR}\sim10^3$ K, which is below the dust sublimation temperature of $T_{\rm sub}\sim1800$ K \citep{2016ApJ...829...19V}.

We neglect the contribution of X-rays to the dust echo as was done in \cite{2023ApJ...948...42W} since the X-ray emission was {first observed more than 300 days after the OUV peak by Swift X-ray Telescope (XRT) and the unabsorbed luminosity was inferred to be $L_X\sim1.5\times10^{45}~{\rm erg~s^{-1}}\ll L_{\rm OUV}$ \citep{2023MNRAS.522.3992W}.} 

{To fit the bolometric IR luminosity, we model $ L_{\rm IR}$ as the convolution of $L_{\rm OUV}$
 with a (normalized) time spreading function $f(t)$, which depends on the spatial distribution of the surrounding  dust \citep{2022PhRvL.128v1101R,2023ApJ...948...42W}: 
\begin{equation}
    L_{\rm IR}(t)=\epsilon_{\rm dust}\epsilon_{\Omega}\int_{-\infty}^{+\infty}L_{\rm OUV}(t')f(t-t')dt',
    \label{eq:L_IR}
\end{equation}
where $\epsilon_{\rm dust}<1$  represents the fraction of the incident radiation that is re-processed to IR radiation by the illuminated dust, and $\epsilon_\Omega$ is the solid angle coverage factor of the dust distribution. Our chosen form of $f(t)$ is inspired from the IR lightcurve in Fig. \ref{fig:4TDEs}.  We notice that, unlike AT2019dsg/fdr/aalc, whose IR emissions are very weak before the OUV peak, for AT2021lwx the IR light curve 
seems to have a component that evolves like $L_{\rm OUV}$ at early times. The late time evolutions of $L_{\rm IR}$ and $L_{\rm OUV}$ suggest that the former may eventually overtake the latter, and persist over a longer time scale. 
%aligns well with the OUV light curve before $t_{\rm pk}$, as shown in the lower panel of Fig. \ref{fig:4TDEs}. 
{These considerations lead to a two-component model of the echo, where the early-time component is attributed to either anisotropic dust distribution or the pre-existing dust around the SMBH (with no time delay with respect to $L_{\rm OUV}$)}, whereas the late-time part is attributed to a dust torus similar to those of AT2019dsg/fdr/aalc. 
%
%component, e.g., $f_{\rm LoS}=\delta(t)$, in addition to the uniform box function that represents a dust torus with a shape of spherical segment, e.g., $f_{\rm S}(t)=1/(2\Delta T_{\rm IR})$ if $0\leq t\leq2\Delta T_{\rm IR}$ otherwise $f_{\rm S}(t)=0$. In the expression of $f_{\rm S}$, $\Delta T_{\rm IR}$ is the time spread of IR emissions. 
}

{Fig. \ref{fig:schematic} schematically illustrates the physical picture of the spherical and early-time components of the dust echo, where the SMBH, accretion disk, disk corona, isotropic wind envelope, dust torus and potentially a jet are shown. The red dashed lines indicate the optical paths that cause the the time delay of the dust torus component due to the extended dust torus. 
Observationally, $2\Delta T_{\rm IR}$ is comparable with the IR time delay defined as the time difference between the IR and OUV peaks in the SMBH frame, with which we can infer the radius of the inner edge of dust torus, i.e., $R_{\rm dust}\simeq c\Delta T_{\rm IR}.$  An external dust cloud, responsible for the undelayed component of $L_{\rm IR}$, is also shown in the figure. Formally, in the function $f(t)$, the early-time component is represented by a Dirac Delta, $f_{\rm ET}(t)=\delta(t)$, whereas the torus component can be represented by the commonly used box function \citep[see, e.g.,][]{2022PhRvL.128v1101R}: $f_{\rm S}(t)=1/(2\Delta T_{\rm IR})$ if $0\leq t\leq2\Delta T_{\rm IR}$ otherwise $f_{\rm S}(t)=0$.}
%as 
%
%In addition, the external cloud/dust along the line of sight would lead to the $f_{\rm LoS}$ term in Eq. \ref{eq:box_func} represented by a Dirac $\delta$ function and result in the LoS IR component arriving simultaneously with OUV emissions.} 
Combining $f_{\rm ET}$ and $f_{\rm S}$, we explicitly write down the normalized time spreading function 
\begin{equation}\begin{split}   
    f(t)
    =f_{\rm ET}+f_{\rm S}=\lambda\delta(t)+\frac{(1-\lambda)}{2\Delta T_{\rm IR}}H(t,0,2\Delta T_{\rm IR}),
    \end{split}
    \label{eq:box_func}
\end{equation}
where the weighting parameter $0\leq\lambda\leq1$ {represents the fraction of total IR power that can be attributed to LoS dust, and $H(x,a,b)$ is the step function, e.g., $H(x,a,b)=1$ if $a<x<b$ otherwise $H(x,a,b)=0$. } 

{We assume an overall dust echo efficiency $\epsilon_{\rm dust}\epsilon_\Omega\simeq0.3-0.4$ comparable to \cite{2023ApJ...948...42W} and use Eqs.~\ref{eq:L_IR} and \ref{eq:box_func} to explain the bolometric IR light curve. Combining $\epsilon_{\rm dust}\epsilon_\Omega$ and ET weighting factor $\lambda$, we infer the dust echo efficiencies for the early-time component and the spherical dust torus component respectively as $\lambda\epsilon_{\rm dust}\epsilon_\Omega$ and $(1-\lambda)\epsilon_{\rm dust}\epsilon_\Omega$.} {The best fit values of $\lambda$ and $\Delta T_{\rm IR}$ are given in Table \ref{tab:params}, whereas the best-fit IR light curves are shown in the upper panel of Fig. \ref{fig:4TDEs}}. Since the IR data is only available up to 300 days after the OUV peak and the IR light curve maintains a flat shape until the latest data point, the IR time delay is therefore uncertain, which is illustrated by the magenta shaded area. The solid magenta line predicts immediate decrease after the latest data point, e.g., $\Delta T_{\rm IR}=180$ d, while the dashed-dotted magenta line corresponds to a more extended dust torus $R_{\rm dust}\sim10^{18}~\rm cm$ as reported in \cite{2023MNRAS.522.3992W}. Table \ref{tab:params} lists the dust echo parameters from the IR interpretation. The dust radius is estimated to be $R_{\rm dust}=c\Delta T_{\rm IR}\sim5.4\times10^{17}-10^{18}~\rm cm$ and is consistent with the
dust sublimation radius \citep[e.g.,][]{2016MNRAS.460..980N,2019ApJ...871...15J}
\begin{equation}
R_{\rm sub}\simeq6.3\times10^{17}~{\rm cm}~
L_{\rm IR, 45.5}^{1/2}T_{\rm sub, 3.25}^{-2.8}a_{\rm dust, -5}^{-0.51},
\end{equation}
where $L_{\rm IR,45.5}=L_{\rm IR}/(10^{45.5}~\rm erg~s^{-1})$, $T_{\rm sub}\simeq1800$ K is the sublimation temperature and $a_{\rm dust}\sim10^{-5}a_{\rm dust, -5}~\rm cm$ is the dust grain radius. 

{Fig. \ref{fig:4TDEs} demonstrates that our two-component model can explain the IR observations very well and could be tested by further IR follow-ups. However, we need to note that based on the current observations, our fitting suggests that a static spherical dust torus is not sufficient to explain the whole IR light curve, and we cannot exclusively determine the physical meaning of the early-time component. For instance, it could arise from the anisotropic or irregular dust distribution, or from pre-existing dust clumps around the SMBH \citep[e.g.,][]{2019ApJ...871...15J}, as shown in Fig. \ref{fig:schematic}. Another possibility is an expanding dust torus pushed by radiation or winds, which introduces the time evolution of the dust compared to a static distribution described by $f_{\rm S}$. A detailed study of the physical interpretation of $f_{\rm ET}$ is beyond the scope of this work. In the following text, we use our two-component fitting to describe the evolution of IR target photons for neutrino production. Our IR light curve can be considered as an ``effective" description that reproduces the data well. Different physical scenarios -- if they fit the data well -- should give a similar light curve. Therefore, our IR model is sufficient for the purpose of this study.}

Using the dust echo efficiency $\epsilon_{\rm dust}\epsilon_\Omega$ in Table \ref{tab:params}, we can estimate the IR-corrected OUV bolometric energy $\mathcal E_{\rm OUV}\approx\mathcal E_{\rm IR}/(\epsilon_{\rm dust}\epsilon_\Omega)\sim0.3M_\odot c^2\simeq\int L_{\rm OUV}dt$, where $\mathcal E_{\rm IR}\sim0.1-0.13 M_\odot c^2$ is the IR bolometric energy obtained by integrating $L_{\rm IR}$ over time. {From this chain of equations, and assuming that the absorbed and unabsorbed OUV luminosities have the same time dependence, we finally obtain $L_{\rm OUV}$. }   One caveat in our IR interpretation is the assumption of $\epsilon_{\rm dust}\epsilon_\Omega$, which renders the unabsorbed $L_{\rm OUV}$ model-dependent. Nevertheless, we will demonstrate later that the IR photons would dominate the neutrino and EM cascade emissions, and our conclusions do not depend sensitively on $\epsilon_{\rm dust}\epsilon_\Omega$.

\begin{deluxetable}{ccchlDlc}[t]
\tablenum{1}
\tablecaption{Physical Parameters for AT2021lwx \label{tab:params}}
\tablewidth{0pt}
\tablehead{
\colhead{\textbf{Description}} &\colhead{\textbf{Parameter}} & \colhead{\textbf{Value}} & \nocolhead{Common} & 
}
\startdata
    {Redshift} &$z$ & 0.995 \\
 {OUV peak time (MJD)} & $t_{\rm pk}$ & 59291\\
 {SMBH mass [$M_\odot$]}  & $M_{\rm BH}$ & $10^8$ \\
 {Star mass [$M_\odot$]} & $M_\star$ & $14$\\
 Peak accretion rate &$\dot M_{\rm BH}(t_{\rm pk})$ &40$L_{\rm Edd}/c^2$\\
 %Accreted Mass  & $\int \dot M_{\rm BH}dt$ & $M_*/2$\\
 \hline
 \textbf{Neutrino observation}&{IC220405B} &{} \\
 \hline
 {Detection time [d]} & $t_{\nu}-t_{\rm pk}$ & $\sim$370\\
 {Energy [TeV]} & $E_\nu$ & 106\\
 {Angular deviation [$^\circ$]} & $\Delta\theta$ & $2.6$\\
\hline
 \textbf{Dust echo modeling}&{} &{} \\
 \hline
  {Dust echo efficiency} & {$\epsilon_{\Omega}\epsilon_{\rm dust}$} &{0.32 (0.43)}\\
  {ET weighting} &{$\lambda$} & {$0.4 ~(0.3)$}\\
    {IR time dispersion [d]} & $\Delta T_{\rm IR}$& $180~(330)$ \\
 {Dust torus radius [cm]} & $R_{\rm dust}$ & $5.4\times10^{17}~(10^{18})$\\
 %{OUV energy} & $\int L_{\rm OUV}dt$ & 0.26 $M_\odot c^2$\\
 %{IR energy} & $\int L_{\rm IR}dt$ & 0.1-0.13$M_\odot c^2$\\
  {Proton efficiency} & $\varepsilon_p$&0.2\\
 {Max proton energy [GeV]} &$E_{p,\rm max}$ & $1.5\times10^{9}$\\
 {Magnetic field [G]} & $B$ & 0.1\\
 \hline
\enddata
\tablecomments{The numbers in parentheses correspond to the case of longer time delay $\Delta T_{\rm IR}=330$ d (magenta dashed line in Fig. \ref{fig:4TDEs}). Data references: $z$, $t_{\rm pk}$, $M_{\rm BH}$ and $M_{\star}$ \citep{2023ApJ...948L..19S}; $t_{\nu}$ and $E_\nu$ \citep{2022GCN.31842....1N}. The dust echo efficiency $\epsilon_\Omega\epsilon_{\rm dust}$ is assumed to be compatible with \cite{2023ApJ...948...42W}.}
\end{deluxetable}

\section{Neutrino and EM cascade emissions}\label{sec:Neu_em}

\begin{figure*}
    \centering
    \includegraphics[width=0.49\textwidth]{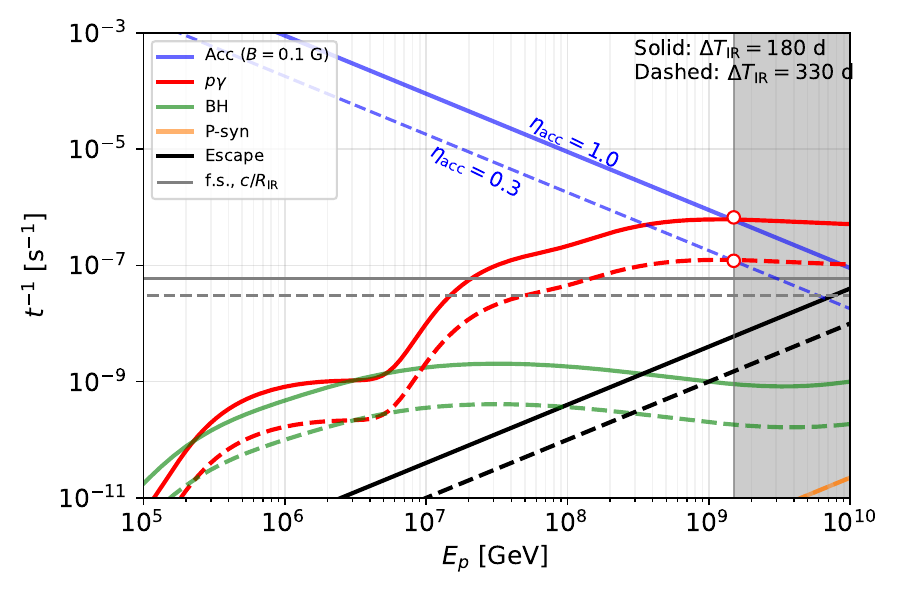}
    \includegraphics[width=0.49\textwidth]{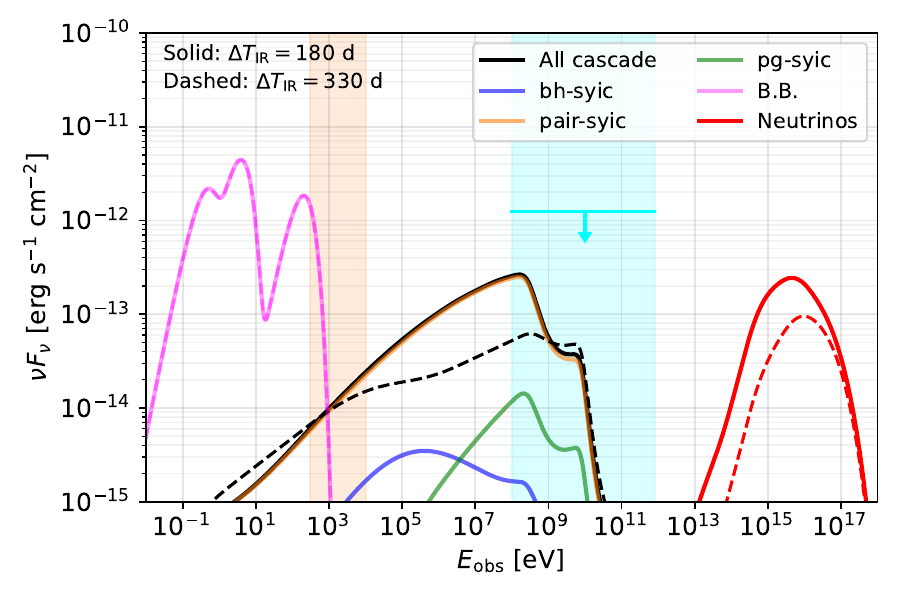}
    \caption{\emph{Left panel:} In-source interaction rates shown at the neutrino detection time $t_\nu$. The solid and dashed curves correspond to the shorter ($\Delta T_{\rm IR}=180$ d) and longer ($\Delta T_{\rm IR}=330$ d) IR time delays with different assumptions for $R_{\rm dust}$, respectively. The gray shaded area shows the region beyond the maximum proton energy, whereas the acceleration efficiencies, $\eta_{\rm acc}=1.0-0.3$ can be used to describe $E_{p,\rm max}$, see red circles. \emph{Right panel:} SED of the muon neutrino (red curves) and EM cascade (black curves) emissions at the neutrino detection time. The magenta curve shows the spectrum of target black body photons. The dashed and solid curves have the same meaning with the left panel. For the $\Delta T=180$ d case, the orange, green and blue dashed curves represent the components of EM cascades. The orange and cyan areas depict the XRT and \emph{Fermi} LAT energy ranges, whereas the \emph{Fermi} upper limit is shown as the cyan arrow.}
    \label{fig:spec}
\end{figure*}

The observational parameters for AT2021lwx and the potentially associated neutrino event IC220405B are summarized in Table \ref{tab:params}. We follow the treatments in \cite{2023ApJ...948...42W} and \cite{2023ApJ...956...30Y} and assume the injected proton luminosity is a fraction of the accretion power, e.g., $L_p=\varepsilon_p\dot M_{\rm BH}c^2$, where an efficient proton injection efficiency $\varepsilon_p=0.2$ is used as the fiducial value as in \cite{2023ApJ...948...42W} for AT2019dsg/fdr/aalc. We assume the accretion rate aligns with the OUV light curve, e.g., $\dot M_{\rm BH}\propto L_{\rm OUV}$ as it is consistent with the $t^{-5/3}$ prediction which reflects the accretion history. The peak accretion rate\footnote{Note that in \cite{2023ApJ...948...42W}, the highly super-Eddington peak accretion rate, $\dot M_{\rm BH}(t_{\rm pk})=100L_{\rm Edd}/c^2$, is used.} $\dot M_{\rm BH}(t_{\rm pk})=40L_{\rm Edd}/c^2$ is estimated to avoid exceeding the hard up limit of the accreted mass $\int\dot M_{\rm BH}dt\lesssim M_\star/2$, where $L_{\rm Edd}=1.3\times10^{46}~{\rm erg~s^{-1}}(M_{\rm BH}/10^8M_\odot)$ is the Eddington luminosity. On the other hand, the peak accretion rate can be interpreted as the super-Eddington accretion with $\dot M_{\rm BH}(t_{\rm pk})/\dot M_{\rm Edd}=\mathcal O(1.0)$, where the Eddington accretion rate $\dot M_{\rm Edd}=L_{\rm Edd}/(\eta_{\rm rad}c^2)$ indicates the accretion rate to power Eddington radiation with the radiation efficiency $\eta_w\sim0.01-0.1$ \citep{2015MNRAS.454L...6M}. Our IR-corrected OUV bolometric peak luminosity reaches $\sim1.2\times10^{46}~\rm erg~s^{-1}$ and is comparable with the Eddington luminosity $L_{\rm Edd}=1.3\times10^{46}~\rm erg~s^{-1}$ for a SMBH of mass $M_{\rm BH}=10^8~M_\odot$. In this case, the peak mass accretion rate can reach a few $\times10 \, L_{\rm Edd}/c^2$ with $\dot M_{\rm BH}(t_{\rm pk})/\dot M_{\rm Edd}=1$ and $\eta_{\rm rad}\sim 0.01-0.1.$ Hence, our fiducial $\dot M_{\rm BH}(t_{\rm pk})=40L_{\rm Edd}/c^2$ is not too optimistic.

We assume a power-law injection rate for the accelerated protons in the isotropic wind region inside the dust torus, e.g., $Q_p\propto E_p^{-2}\exp(-E_p/E_{p,\rm max})$ and normalize the spectrum with $\int E\dot Q_pdE_p=L_p/V$, where $V\approx4\pi R_{\rm dust}^3/3$ is the volume within the dust\footnote{In \cite{2023ApJ...956...30Y}, the more compact radiation zones, where thermal OUV photons dominate the $p\gamma$ interactions, are discussed as well. However, for AT2019lwx, a smaller radius ($10^{16}-10^{17}$ cm) would lead to very bright electromagnetic cascade emission which would contradict the non-detection of $\gamma$-rays by \emph{Fermi} LAT; this scenario is therefore omitted.}. Without explicitly specifying the accelerator, we instead parameterize the acceleration zone by the maximal proton energy $E_{p,\rm max}$. In general, the protons can be energized in the compact inner jet, accretion disk or disk corona, or an extended isotropic wind \citep[e.g.,][]{2020ApJ...902..108M} with magnetic field strength comparable to AGNs, e.g., $B\sim0.1-1$ G. 

While propagating inside the radiation zone (the yellow region in Fig. \ref{fig:schematic}), the protons will undergo photomeson ($p\gamma$) and hadronuclear ($pp$) energy losses via interactions respectively with target thermal photons and wind protons. The resulting neutral ($\pi^0$) and charged ($\pi^\pm$) pions would decay into neutrinos, $\gamma$ rays and secondary electrons. These secondary electrons together with the electron/positron pairs generated from $\gamma\gamma$ annihilation and Bethe-Heitler (BH) interactions will subsequent initiate EM cascade emissions via synchrotron and inverse Compton radiation. {We denote the EM cascade components originating from $p\gamma$, $\gamma\gamma$, and BH processes respectively as `pg-syic', `pair-syic', and `bh-syic'.} To obtain the neutrino and EM cascade spectra, we use AM$^3$ \citep[Astrophysical Multi-Messenger Modeling,][]{2017ApJ...843..109G,2023arXiv231213371K} software to numerically solve the coupled time-dependent transport equations for all relevant particle species; see \cite{2023ApJ...956...30Y} for a detailed description and discussion of the transport equations, including the particle injection, energy loss and escape terms. 

\begin{figure*}
    \centering
    \includegraphics[width=0.49\textwidth]{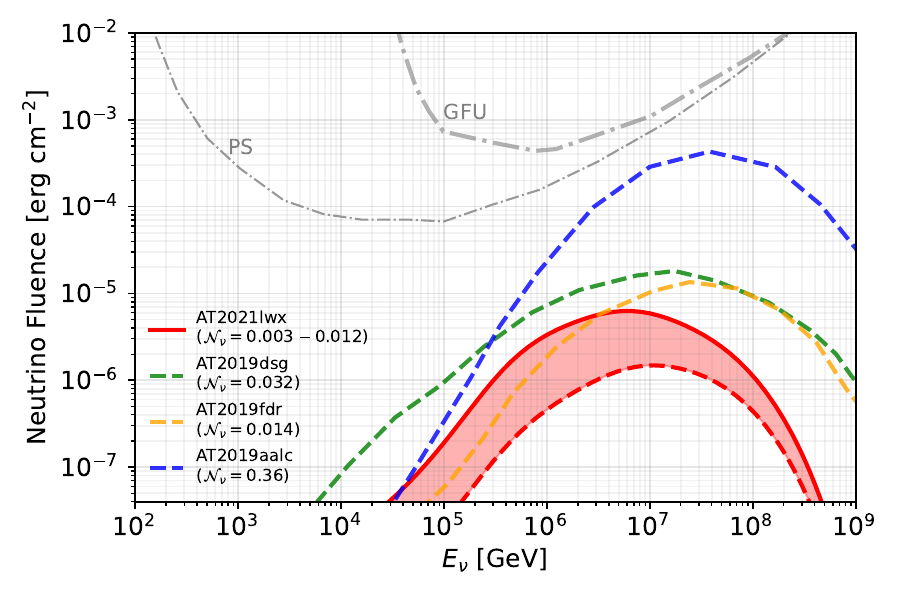}
    \includegraphics[width=0.49\textwidth]{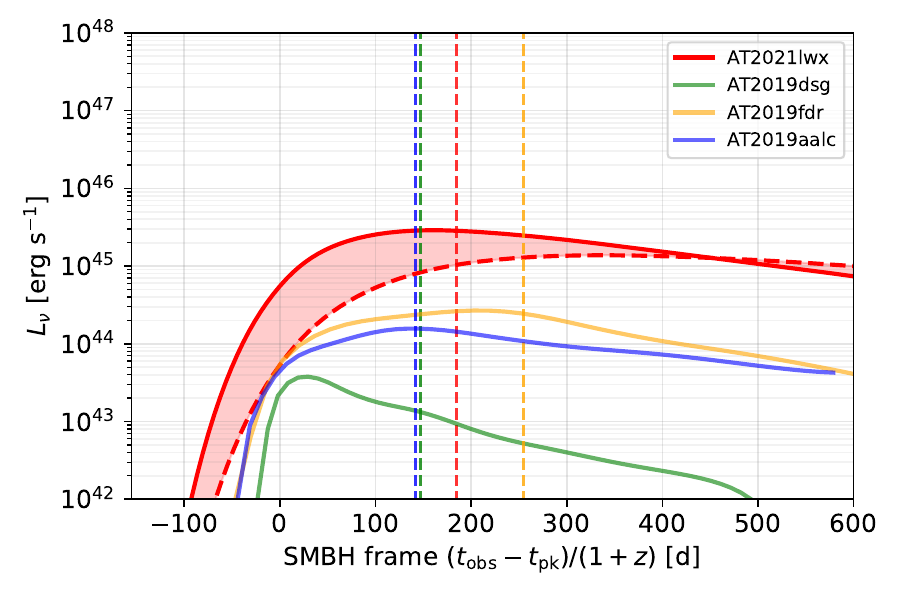}

    \caption{\emph{Left panel:} Cumulative single-flavor neutrino fluences at $t_\nu$ for AT2021lwx (red curves) and the other three TDEs, AT2019dsg/fdr/aalc \cite[green/orange/blue dashed curves, taken from][]{2023ApJ...948...42W}. The thin and thick dashed-dotted gray curves show the IceCube sensitivities for point-source and GFU searches. {The uncertainties in IR lightcurve interpretation leads to the expected GFU neutrino number in the range $\mathcal N_\nu=3\times10^{-3}-0.012$.} \emph{Right panel:} Neutrino luminosities for AT2021lwx and the other three TDEs measured in the SMBH frame. The vertical lines represent the corresponding neutrino detection times. %In both panels, only the spherical component (e.g., the magenta curves in Fig. \ref{fig:4TDEs}) of IR photons are taken into account, 
    {The solid (dashed) curves correspond to $\Delta T_{\rm IR}=$ 180 d (330 d), and the red areas correspond to the uncertainties from dust echo interpretations.} }
    \label{fig:NeuFluence}
\end{figure*}

{As demonstrated by \cite{2023ApJ...948...42W} and \cite{2023ApJ...956...30Y}, the contribution to neutrino and EM cascade emissions from $pp$ interactions is typically subdominant compared to the $p\gamma$ processes, even if a significant fraction of the accreted mass is converted to non-relativistic winds with velocities of $0.1c$ \citep{2018ApJ...859L..20D,2020PhRvD.102h3013Y,2021ApJ...911L..15Y}. In the following text, we specifically focus on the $p\gamma$ contributions. For the target photon fields, we consider the thermal IR, OUV and X-ray photons isotropized within the dust radius. {Since the early-time (ET) component in the IR light curve interpretation could be produced by the dust outside the radiation zone, as shown in Fig. \ref{fig:schematic}, we consider only the dust torus component for the IR target photons, i.e., the magenta curves in the upper panel of Fig. \ref{fig:4TDEs}, in a conservative case.} For the X-ray component, we assume a constant luminosity of $L_X=1.5\times10^{45}~\rm erg~s^{-1}$ as in \cite{2023ApJ...948...42W} and use an AT2019dsg-like temperature $k_BT_X=72$ eV. We use the magnetic field strength $B=0.1$ G as the fiducial value as in \cite{2023ApJ...948...42W}. }

Fig. \ref{fig:spec} shows the proton energy loss rates (left panel) and the neutrino/EM cascade spectral energy distribution (SEDs, right panel) produced in the isotropic wind of radius $R_{\rm dust}$ at neutrino detection time $t_\nu$. We assume the maximum energy of the injected proton to be $1.5\times10^9$ GeV. In the left panel, the red green, orange, and black curves respectively depict the $p\gamma$, BH, proton synchrotron and escape rates. The horizontal gray lines show the free streaming rates ($t_{\rm f.s.}^{-1}=c/R_{\rm dust}$) for neutral particles. The cases of shorter and longer IR time dispersion, e.g., $\Delta T_{\rm IR}=$180 d and 330 d, are shown as the dashed and solid curves. {In both cases, the $p\gamma$ interactions are efficient and fast in the proton energy range $E_{p}\sim10^{7}-10^9$ GeV, e.g., $t_{p\gamma}^{-1}/t_{\rm f.s.}^{-1}>1$,} which implies that the neutrino radiation is mainly constrained by the proton luminosity. Given the acceleration efficiency $\eta_{\rm acc}\lesssim1$, the proton acceleration rates, $t_{p,\rm acc}^{-1}=\eta_{\rm acc}eBc/E_p$, are illustrated as blue lines. Despite $E_{p,\rm max}$ is treated as a free parameter without specifying the acceleration sites, we demonstrate that $E_{p,\rm max}=1.5\times10^{9}~\rm GeV$ is achievable within the wind {(one possible site for proton acceleration)} for the magnetic field $B=0.1$ G. The red circles in the left panel demonstrate that the chosen $E_{p,\rm max}$ can be obtained for reasonable/conservative acceleration efficiencies $\eta_{\rm acc}=0.3$ and $1.0$ for the IR time spreads $\Delta T_{\rm IR}=330$ d and 180 d, respectively, from balancing the acceleration rate with the proton interaction rate.

The right panel of Fig. \ref{fig:spec} shows the SEDs for target photons (magenta curves), overall EM cascades (black curves) and neutrinos (red curves). The dashed and solid curves have the same meaning with the left panel. For the shorter IR time delay case ($\Delta T_{\rm IR}=180$ d), the orange, green and blue curves  illustrate the contributions from secondary electrons/positions originated from $\gamma\gamma$ annihilation, $p\gamma$ and BH processes, respectively. {The $\gamma$-rays from $\pi^{0}$ decays are completely depleted via $\gamma\gamma$ annihilation with in-source thermal photons and extragalactic background lights.} A more detailed and quantitative description of the EM cascade SEDs can be found in \cite{2023ApJ...956...30Y}. The orange and cyan areas depict the Swift XRT and \emph{Fermi} Large Area Telescope (LAT) energy ranges. The non-detection of $\gamma$-rays by \emph{Fermi} LAT in the direction of IC220405B place an upper limits in the energy range $0.1-800$ GeV \citep{2022GCN.31845....1G}, shown as the cyan upper limit. We find that our results are consistent with the observational constraints even for the optimistic parameter sets.

The comparison of the predicted neutrino fluence, obtained by integrating the flux, and neutrino luminosity in the SMBH frame of AT2021lwx with AT2019dsg/fdr/aalc are illustrated in the left and right panels of Fig. \ref{fig:NeuFluence}. The red areas correspond to the uncertainties from IR interpretations, e.g., $\Delta T_{\rm IR}=180-330$ d and $R_{\rm dust}=5.4\times10^{17}-10^{18}~\rm cm$. The IceCube sensitivities for point source \citep[PS,][]{2014ApJ...796..109A} and gamma-ray follow up \citep[GFU,][]{2019ICRC...36.1021B} searches are ploted in the thin gray dashed-dotted curves. We observe that the neutrino spectra of AT2021lwx are similar to those of the other three TDEs but at a lower fluence level due to the high redshift. Using the GFU effective area\footnote{The GFU neutrino numbers, estimated using IceCube GFU effective areas \citep{2016JInst..1111009I}, are more suitable when comparing the model predictions to actual follow-up observations, whereas point source (PS) neutrino numbers are typically used for independent point-source analyses.},  we estimate the expected neutrino number from AT2021lwx to be in the range of $\mathcal N_\nu\simeq3.0\times10^{-3}-0.012$, which is lower than the other three TDEs. From the right panel, the peak time of the neutrino luminosity of AT2021lwx, e.g., 100-200 days in the SMBH frame, could explain the time delay of IC220405B (vertical red dashed line). 
%Moreover, the neutrino light curves of these four TDEs exhibit similar luminosities and time delays. For instance, the peak luminosities fall roughly in the range of $L_{\nu}\sim10^{44}-10^{45}~\rm erg~ s^{-1}$, and the neutrino time delays are distributed in the range of 150-300 days in the SMBH frame. This further consolidates the assumption that all neutrino-coincident TDE candidates could be powered by similar physical processes. 

\section{Discussion}\label{sec:dission}

Let us first of all note that the description of the IR emission is a crucial part in our scenario, because the chosen maximal proton energies reaching $E_{p,\rm max}\simeq10^9$ GeV allow for $p\gamma$ interactions beyond the threshold with the abundant IR photons -- which are dominating the  neutrino and accompanying EM cascade emissions. Moreover, the radius of the dust torus, which determines the target photon density and consequently the $p\gamma$ interaction efficiency,
can be inferred from the time delay appeared in the IR light curves, defined as the time difference between OUV and IR peaks. We propose an interpretation of the IR light curve that consists of a spherical dust torus and an early-time components, denoted as $f_{\rm S}$ and $f_{\rm ET}$,  respectively. 

The primary uncertainty in the multi-messenger modeling of AT2021lwx arises from the lack of IR data since 300 days after the OUV peak in the SMBH frame, which leads to the uncertainties in the evaluation of the time delay, the time dispersion $\Delta T_{\rm IR}$ of the dust torus component (the magenta curves in Fig. \ref{fig:4TDEs} and the $f_{\rm S}$ term in Eq. \ref{eq:box_func}), and equivalently the radius $R_{\rm dust}=c\Delta T_{\rm IR}$. Our IR interpretation demonstrates that the time dispersion ($\sim$ half of the time delay in the SMBH frame) lies in the range $\Delta T_{\rm IR}\simeq180-330$ d, which corresponds to the dust radius $R_{\rm dust}=c\Delta T_{\rm IR}\simeq(5.4-10)\times10^{17}$ cm. The resulting uncertainties in neutrino and EM emissions are illustrated in Figs.~\ref{fig:spec} and~\ref{fig:NeuFluence}, and the predicted neutrino number is limited to be $\mathcal N_\nu\simeq3\times10^{-3}-0.12$. Further follow-up observations up to $2(1+z)\Delta T_{\rm IR}\sim1300$ days after the OUV peak are advisable to obtain more stringent constraints on the neutrino number and on our model.

{On the other hand, two-component dust echo scenario is constructed to interpret the early-time IR light curve. There could be other alternative models, such as the dust clumps in the broad-line regions for TDEs in AGNs \citep{2019ApJ...871...15J} or the bounded/unbounded debris. If the early IR emissions are produced within the dust radius (i.e., our radiation zone), one should take these IR photons (inferred from the cyan curve in the upper panel of Fig. \ref{fig:4TDEs}) into account as additional targets for the $p\gamma$ interactions. We tested that potential additional contribution of early time IR photons and found that the neutrino fluence is affected by a factor less than 1.5. The reason is that the system is already $p\gamma$ efficient, and the neutrino power is limited by the injected proton luminosity, which is determined by $\varepsilon_p\dot M_{\rm BH}(t_{\rm pk})$ and is also constrained by the \emph{Fermi}-LAT upper limit.}

Our multi-messenger model, which gives $\mathcal N_\nu\lesssim0.012$, seems to  disfavor the neutrino-TDE coincidence together with the misalignment of the TDE outside the neutrino 90\% error box. However, aside from the similarities with AT2019dsg/fdr/aalc and the potential correlation with IC220405B, AT2021lwx remains an important TDE candidate, being one of the non-jetted TDEs with the highest redshifts \citep[see, e.g.,][for a TDE sample]{2023ApJ...955L...6Y}, and could have profound implication on the redshift distribution of TDEs, including the SMBH mass and the mass of disrupted stars. For AT2021lwx, the total energy released in the OUV bands reaches $\mathcal E_{\rm OUV}=\int L_{\rm OUV} dt\sim0.3M_\odot c^2$, which implies an accreted mass $M_{\rm acc}\sim\mathcal E_{\rm OUV}/(\eta_{\rm rad}c^2)\sim3-30~M_\odot$\footnote{Such estimation is based on the energy conversion and does not depend sensitively on the classification of AT2021lwx.} given the radiation efficiency $\eta_{\rm rad}\sim0.1-0.01$. Indeed, the volumetric rates, i.e.,  $\dot\rho_{\rm TDE,\star}$, of  TDEs with $M_{\rm BH}\sim10^8~M_\odot$ and $M_{\rm acc}\sim3-30M_{\odot}$ are low. On the other hand, given the high redshift of AT2021lwx, such detection is not impossible especially for a high cosmological volume if the event is bright enough. Considering a rapid redshift evolution, e.g., $\dot\rho_{\rm TDE,\star}\propto(1+z)^{-3}$, we compare  the relative TDE rate, i.e., $\dot N_{\rm TDE}(z<z_{\rm lim})\sim V_{\rm co}(z_{\rm lim})\dot\rho_{\rm TDE,\star}(z_{\rm lim})$, within redshift $z_{\rm lim}=0.995$ to the AT20l9fdr-like redshift $z_{\rm lim}=0.26$, $\dot N_{\rm TDE}(z<0.995)/\dot N_{\rm TDE}(z<0.26)\sim V_{\rm co}(0.995)/[4V_{\rm co}(0.26)]\sim8$, where $V_{\rm co}(z)$ is cosmological comoving volume at $z$. This implies that a powerful object involving a massive star has a larger abundance across large cosmological volumes. Similar results could be obtained using the TDE rate inferred from star formation history and SMBH mass function \citep[e.g.,][]{2016MNRAS.461..371K}.

In addition to the IR photons, $p\gamma$ interactions with X-ray photons could also dominate neutrino production within a relatively compact radiation zone, as proposed by \cite{2023ApJ...948...42W} (model ``M-X''); this option is attractive because it requires much lower maximal proton energies, and, thus, a much less efficient accelerator. For AT2021lwx, due to the absence of early-time X-ray observations and incomplete information about the radius of the X-ray emitters, we focus on the dust echo model and do not consider the scenario where X-rays are dominant.

\section{Summary and conclusions}

We have investigated the potential correlation between the neutrino event IC220405B and an energetic TDE candidate, AT2021lwx, at redshift $z=0.995$. In addition to luminous thermal OUV emissions, AT2021lwx exhibits bright and long-lasting IR luminosities, which can be interpreted as a strong dust echo incorporating the dust torus  component originating from the dust torus and the early-time contributions. We have pointed out that AT2021lwx shares crucial similarities with the other three neutrino-coincident TDEs and TDE candidates, e.g., AT2019dsg/fdr/aalc, including strong X-ray/OUV emissions, strong dust echoes, and comparable time delays ($\sim150-300$ days) of the neutrino detection with respect to the OUV peaks in the SMBH rest frame. We have studied the neutrino and EM cascade emissions from an isotropic radiation zone inside the dust radius in a fully time-dependent manner following the treatments in \citet{2023ApJ...948...42W} and \citet{2023ApJ...956...30Y}. We have demonstrated that the outputs, such as the EM/neutrino SEDs, neutrino light curves and fluences, are qualitatively consistent with the other three TDEs. Especially, the neutrino time delay could be explained by dust echo target photons. These similarities make AT2021lwx an interesting target and imply that these objects may share similar underlying physical processes.

Our results indicate that, in consistency with the non-detection of $\gamma$-rays by \emph{Fermi}-LAT, the expected neutrino numerical number is limited to the range $\mathcal N_\nu\simeq3.0\times10^{-3}-0.012$, which is expected for far away sources and might suggest that the association could have low significance. However, the expected event rate is not very different from the other three TDEs, and such low event rates are expected for single neutrino event detections from many far-away sources \citep[Eddington bias, see][]{2019A&A...622L...9S}. 

We suggest conducting further multi-wavelength follow-ups, especially in the IR band, on this object for an extended period. Additionally, we recommend studying the neutrino track reconstruction in this particular case for a more definitive conclusion regarding the neutrino correlation.
{The extended IR observations, such as the upcoming annual data release of the WISE survey, along with the confirmation or exclusion of the neutrino coincidence, would test our dust echo model and shed more light on the physical picture of TDEs, such as the geometry of the dust torus and the origin of the early time IR component. After all, our time-dependent multi-messenger diagnosis, consisting of the neutrino and EM cascade counterparts, provides a comprehensive and generic template for interpreting the spectral and temporal signatures of future neutrino-coincident TDEs.
\\

{Note added: after the paper was submitted, the NEOWISE 2024 data was released, extending the IR light curve to MJD 60252, equivalent to approximately 481 days after the OUV peak in the SMBH rest frame. The latest data release\footnote{The data is available at \url{https://irsa.ipac.caltech.edu/frontpage/}} indicates that the late-time W1/W2 apparent magnitudes remained roughly unchanged from the early epochs. We infer that the unbinned data points lie within the uncertainties of our model (red shaded areas in Fig. \ref{fig:4TDEs}). Refined analyses of the late-time IR data would be needed to obtain a robust constraint on $R_{\rm IR}$.}}

%\section{Summary and conclusions}\label{sec:summary}

\acknowledgments

We would like to thank Simeon Reusch, Marek Kowalski, and Ning Jiang for useful discussions, and Xin-Yue Shi for thorough internal review.  CL acknowledges support from the NSF grant PHY-2309973, and from the National Astronomical Observatory of Japan, where part of this work was conducted. 

\bibliography{ref.bib}
%\appendix 
%\section{Proton interaction rates}\label{app:p_rates}
%\begin{figure}
%    \includegraphics[width=0.49\textwidth]{Figs/P_rates.pdf}
%    \caption{Interaction rates}
%    \label{fig:p_rates}
%\end{figure}
\end{CJK*}
\end{document}